\documentclass[aip,jap,superscriptaddress,reprint,twocolumn]{revtex4}%
\usepackage{grffile}
\usepackage{epstopdf}
\usepackage{graphicx}
\usepackage{endnotes}
\usepackage{bm}
\usepackage{epsfig}
\usepackage{amsmath}
\usepackage{color}
\usepackage{textcomp}
\usepackage[colorlinks=true,
            linkcolor=blue,
            urlcolor=blue,
            citecolor=blue,
            hyperfootnotes=true]{hyperref}

\newcommand{\kp}{{$k\cdot p$ }}
\newcommand{\ie}{{\it i.e. }}

\newcommand{\LiO}{{LiOsO$_3$ }}
\newcommand{\Rc}{{$R\bar{3}c$ }}
\newcommand{\rc}{{$R3c$ }}

\newcommand{\Est}[1]{\expandafter\hat#1}
\newcommand{\My}{\widetilde{M}_{y} }
\newcommand{\Mi}{\widetilde{M}_{i} }
\newcommand{\Cz}{C_{3z} }
\newcommand{\T}{\mathcal{T} }

\begin{document}

\title{Nonsymmorphic cubic Dirac point and crossed nodal rings across the ferroelectric phase transition in \LiO}

\author{Wing Chi Yu$^*$}
\affiliation{Center for Advanced 2D Materials and Graphene Research Center, National University of Singapore, Singapore 117546, Singapore}

\author{Xiaoting Zhou\footnote{These authors contributed equally to this work.}}
\email{physxtzhou@gmail.com}
\affiliation{Center for Advanced 2D Materials and Graphene Research Center, National University of Singapore, Singapore 117546, Singapore}
\affiliation{Physics Division, National Center for Theoretical Science, Hsinchu 30013, Taiwan}

\author{Feng-Chuan Chuang}
\affiliation{Department of Physics, National Sun Yat-Sen University, Kaohsiung 804, Taiwan}

\author{Shengyuan A. Yang}\email{shengyuan\_yang@sutd.edu.sg}
\affiliation{Research Laboratory for Quantum Materials, Singapore University of Technology and Design, Singapore 487372, Singapore}

\author{Hsin Lin}\email{nilnish@gmail.com}
\affiliation{Center for Advanced 2D Materials and Graphene Research Center, National University of Singapore, Singapore 117546, Singapore}
\affiliation{Department of Physics, National University of Singapore, Singapore 117546, Singapore}
\affiliation{Institute of Physics, Academia Sinica, Taipei 11529, Taiwan}

\author{Arun Bansil}
\affiliation{Department of Physics, Northeastern University, Boston, Massachusetts 02115, USA}


\begin{abstract}
Crystalline symmetries can generate exotic band-crossing features, which can lead to unconventional fermionic excitations with interesting physical properties. We show how a cubic Dirac point---a four-fold-degenerate band-crossing point with cubic dispersion in a plane and a linear dispersion in the third direction---can be stabilized through the presence of a nonsymmorphic glide mirror symmetry in the space group of the crystal. Notably, the cubic Dirac point in our case appears on a threefold axis, even though it has been believed previously that such a point can only appear on a sixfold axis.
We show that a cubic Dirac point involving a threefold axis can be realized close to the Fermi level in the non-ferroelectric phase of LiOsO$_3$. Upon lowering temperature, \LiO has been shown experimentally to undergo a structural phase transition from the non-ferroelectric phase to the ferroelectric phase with spontaneously broken inversion symmetry. Remarkably, we find that the broken symmetry transforms the cubic Dirac point into three mutually-crossed nodal rings. There also exist several linear Dirac points in the low-energy band structure of LiOsO$_3$, each of which is transformed into a single nodal ring across the phase transition.
\end{abstract}


\maketitle

\section{Introduction}
In the past decade, the study of topological phases of matter has become one of the most active areas of research in condensed matter physics \cite{Bansil2016,XLQi2011,Hasan2010,BHYan2017}. Extending the earlier work on gapped topological phases, such as topological insulators~\cite{Kane2005a,Kane2005b,Bernevig2006,LFu2007b,Moore2007,Roy2009,Konig2007,Hsieh2008,Hsieh2009a,Hsieh2009b,Chen2009,Xia2009} and topological superconductors~\cite{Schnyder2008,Kitaev2009,Chung2009,Qi2009a,Zhang2009,Fu2008,Qi2010a,Santos2010,Sau2010,Das2012,SAYang}, the focus has shifted recently to gapless phases, especially the so-called topological semimetals (TSMs)  \cite{Review_hasan2015,Report_balents2011,Review_Ashvin2013,Review_weylburkov2017,YangRW,Review_nlFang2016}.

TSMs are characterized by protected band-crossings in momentum space, which can be zero-dimensional nodal points or one-dimensional nodal lines. Around these band-crossings, electrons behave drastically differently from the conventional Schr\"{o}dinger fermions. For example, the low-energy electrons in three-dimensional (3D) Dirac~\cite{dirac_Young2012,dirac_Wang2012,dirac_Liu12014,dirac_Wang2013,dirac_Chang2017} and Weyl semimetals~\cite{weyl_ashwin2011,weyl_singh2012,weyl_Weng2015,weyl_SMH2015} resemble the relativistic Dirac and Weyl fermions with linear dispersion, allowing investigation of related exotic phenomena, which have previously belonged to the domain of high-energy physics, in a desktop materials setting.  In fact, the family of nodal-line semimetals~\cite{nodal_thBurkov2011,nodal_thAji2014,nodal_thChiu2014,nodal_Fu2015,Mullen2015,nodal_Weng2015,YChen,RYu,YKim,XQChen,SLi,hourl1,hourl2} harbors an even richer topological structure, and supports nodal chains~\cite{nodal_chain,SSWang}, crossing nodal lines~\cite{MZeng,nodal_CLN,XLSheng} as well as Hopf links~\cite{GuoqingHopf,WChen,CZhong,ZYan,PChang,Ezawa}, which have no counterparts in high-energy physics. In addition to these novel bulk fermionic excitations, TSMs also possess exotic topological surface states in the form of Fermi arcs in Dirac/Weyl semimetals and drumhead surface states in nodal-loop semimetals~\cite{weyl_ashwin2011,Review_nlFang2016,SAYang}.

Nonsymmorphic crystalline symmetries, i.e. symmetries which involve a fractional lattice translation, can generate TSMs with exotic type of band crossing features. For example, a glide mirror or a two-fold screw axis can give rise to an hourglass dispersion~\cite{hour}, hourglass loop/chain~\cite{hourl1,hourl2,nodal_chain,SSWang}, and 2D spin-orbit Dirac points~\cite{YK,2DSDP}, which are robust against spin-orbit coupling (SOC) effects.

Here, we show that the glide mirror symmetry, combined with symmorphic three-fold rotation, inversion, and time-reversal symmetries, can generate a cubic Dirac point. This cubic Dirac point is four-fold degenerate with an associated band dispersion, which is cubic in a plane and linear in the third direction normal to the plane. Notably, as a result of the nonsymmorphic symmetry, the cubic Dirac point here appears on a threefold axis, even though it has been believed previously that such a point can only appear on a sixfold axis~\cite{YangNaga,QLiu2017}. We further show that when the inversion symmetry is broken, this cubic Dirac point transforms into three crossed nodal rings.

Through first-principles electronic structure computations, we show that LiOsO$_3$ provides a materials platform for realizing the novel topological physics outlined in the preceding paragraph. \LiO is the first example of a ferroelectric metal~\cite{Anderson1965}. In a recent experiment, Shi \emph{et al.} report that LiOsO$_3$ undergoes a ferroelectric phase transition at a critical temperature $T^*\approx 140$ K~\cite{Shi2013}, at which the material transforms from a centrosymmetric $R\bar{3}c$ structure to a non-centrosymmetric $R3c$ structure, while remaining metallic. \LiO thus realizes the scenario proposed in Ref.~\cite{Anderson1965}. Notably, a related material, HgPbO$_3$, has been suggested recently to host a ferroelectric Weyl semi-metal phase~\cite{RLi}.

Our analysis shows that the non-ferroelectric \Rc structure satisfies the symmetry requirements for hosting a cubic Dirac point. Also, when the inversion symmetry breaks at the ferroelectric transition, we find that the cubic Dirac point gives rise to three crossed nodal rings in the ferroelectric $R3c$ structure. The low-energy band structure of LiOsO$_3$ further shows the presence of several linear Dirac nodes, and we show that these nodes turn into nodal loops during the ferroelectric phase transition. Our study thus not only discovers a new mechanism for generating a cubic Dirac point, but it also offers a promising new materials platform for exploring the interplay between structural transitions, ferroelectricity, and novel topological fermions.

\section{Computational Methods}

Our \emph{ab-initio} calculations were performed by using the all-electron full-potential linearized augmented plane-wave code WIEN2k~\cite{Schwarz2003}. Generalized gradient approximation (GGA) of Perdew, Burke and Ernzerhof (PBE)~\cite{Perdew1996} was employed for the exchange-correlation potential. Experimental lattice constants~\cite{Shi2013} were used in the calculations and the internal structure was optimized. The self-consistent iteration process was repeated until the charge, energy and force converged to less than 0.0001e, 0.00001 Ry, and 1 mRy/a.u, respectively.  SOC was included by using the second variational procedure \cite{MacDonald1980}. The muffin-tin radii were set to 1.47, 1.84, and 1.51 a.u. for the Li, Os, and O atoms, respectively. $R_{\rm{MT}}^{\min}K_{\max}=7$ and a $k$-point mesh of 4000 in the first Brillouin zone were used. An effective tight-binding model was constructed via projection onto the Wannier orbitals~\cite{Mostofi2014,Marzari1997,Souza2001,Marzari2012,Kunes2010} for carrying out topological analysis of the electronic spectrum.We used the Os $d$ orbitals without performing the maximizing localization procedure. Surface states were obtained using the surface Green's function technique~\cite{Sancho2000} applied on a semi-infinite slab.


\section{Crystal Structure and Symmetry}
\begin{figure}[t]
  \centering
  \includegraphics[width=8.5cm]{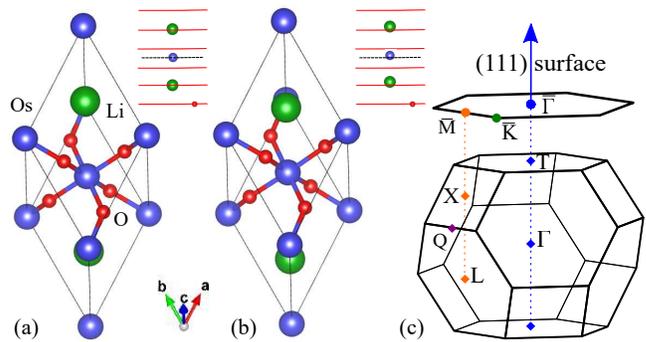}
  \caption{Crystal structures of \LiO  (a) The non-ferroelectric phase and (b) the ferroelectric phase. The insets are schematic drawings of the oxygen planes (red lines) and the location of the Os and Li atoms relative to these planes. (c) Brillouin zone (BZ) of \LiO and the projected (111) surface BZ.}\label{fig1:latticeBZ}
\end{figure}

The non-ferroelectric phase of \LiO has a rhombohedral structure with space group \Rc (No.167) [Fig.~\ref{fig1:latticeBZ}(a)]. The Os atom is located at the center between two Li atoms and also at the symmetric position of two O planes, which preserves inversion symmetry $\mathcal{P}$. Upon lowering the temperature below $T^*$, the two Os atoms are displaced along the [111] direction, resulting in the loss of inversion center [Fig.~\ref{fig1:latticeBZ}(b)]~\cite{Shi2013}. The crystal symmetry is thus lowered to $R3c$ (No.161) in the ferroelectric phase.

We emphasize that the key physics underlying the cubic Dirac point and its transformations in \LiO is controlled by symmetry, and not by the details of the specific material involved. The important symmetries are as follows. The \Rc space group of the non-ferroelectric phase can be generated by three symmetry elements: the inversion $\mathcal{P}$, the three-fold rotation along [111]-direction ($z$-axis) $c_{3z}$, and the glide mirror with respect to the ($\bar{1}10$) plane ($xz$-plane) $\widetilde{M}_y= \lbrace M_{y}|\frac{1}{2} \frac{1}{2} \frac{1}{2} \rbrace$, which is a nonsymmorphic symmetry. In the ferroelectric phase, when $\mathcal{P}$ is broken, the generators of the remaining \rc space group reduce to $c_{3z}$ and $\My$ only. Lattices in both phases are bipartite: there are two sublattices (denoted as A and B) with a relative displacement of $(\frac{1}{2},\frac{1}{2},\frac{1}{2})$. $\mathcal{P}$ and $\My$ map between the two sublattices ($A\leftrightarrow B$), whereas $\Cz$ maps within each sublattice ($A(B)\leftrightarrow A(B))$. The system preserves the time-reversal symmetry $\T$, since no magnetic ordering has been observed in either phase~\cite{Shi2013}.

\section{Band structures without SOC}

\begin{figure}\centering
      \resizebox{8.5cm}{!}{
              \includegraphics{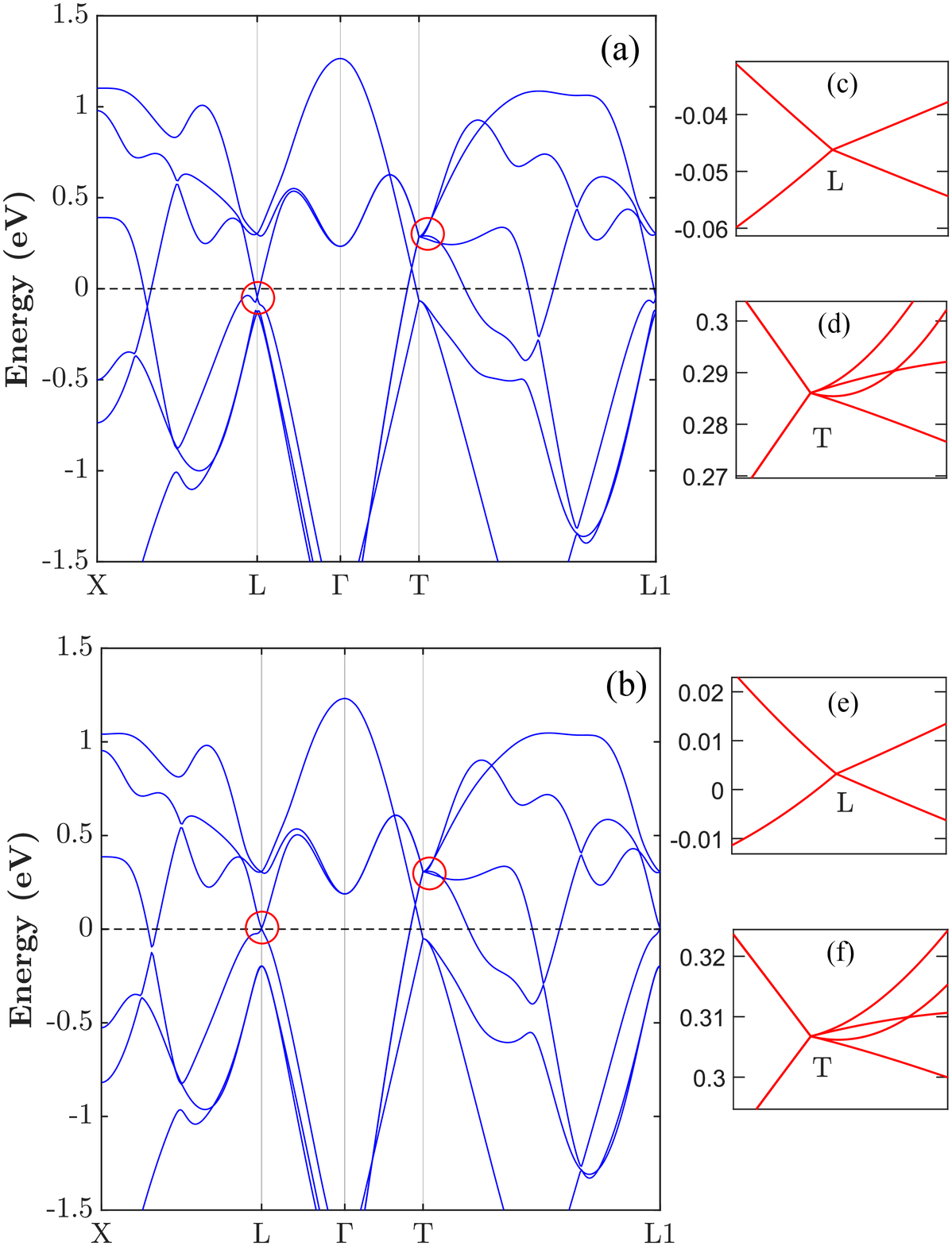}}                				
              \caption{\label{figS1} Electronic band structure of \LiO {in the absence of SOC} for (a) the non-ferroelectric phase and, (b) the ferroelectric phase. (c)-(f) Zoom-in images of the band crossings at the L and T symmetry points marked by red circles in (a) and (b).}
\end{figure}

Without the SOC, the band structures of non-ferroelectric and ferroelectric phases of LiOsO$_3$ are similar as seen in Fig.~\ref{figS1}. Both phases exhibit two-fold degenerate (four-fold if spin-degeneracy is counted) nodal lines passing through the L and T points of the BZ perpendicular to the glide mirror planes. These features of the band structure can be understood by noting that the L and T are time-reversal invariant momentum (TRIM) points, which also reside on the glide mirror planes $\Mi$ ($i=1,2,3$) (the other two glide mirrors are related to $\My$ by $\Cz$), see Fig.~\ref{fig1:latticeBZ}(c). It is easily seen that the line passing through an L point and normal to the corresponding mirror plane $\Mi$ is invariant under the anti-unitary symmetry operation $\mathcal{T}\Mi$, so that
\begin{equation}
(\mathcal{T}\Mi)^2=e^{-ik_z^L}=-1,
\end{equation}
where we have used $\mathcal{T}^2=+1$ for a spinless system and $k_z^L$ is the $k_z$-component of L. The $\mathcal{T}\Mi$ symmetry thus guarantees a two-fold Kramers-like degeneracy on this line. A similar analysis applies to the T point: Since T lies on the intersection of three glide mirrors, three such nodal lines pass through T. Moreover, since $\widetilde{M}_{i+1} = \Cz \widetilde{M}_i \Cz^{-1}$, $[\Cz, \My] \neq 0$, which leads to $[\Mi, \widetilde{M}_{j\neq i}] \neq 0$. For an eigenstate $|\psi_0\rangle$ at T, the three states $|\psi_{i=1,2,3} \rangle= \T \Mi |\psi_{0}\rangle$ must be orthogonal to each other, indicating a four-fold (eight-fold if counting spin) degeneracy at T. This analysis is in accord with our DFT band structure results.

\section{Band structures with SOC}

\subsection{Non-ferroelectric phase: linear and cubic Dirac points}

\begin{figure}[tbh]
  \centering
  \includegraphics[width=8.5cm]{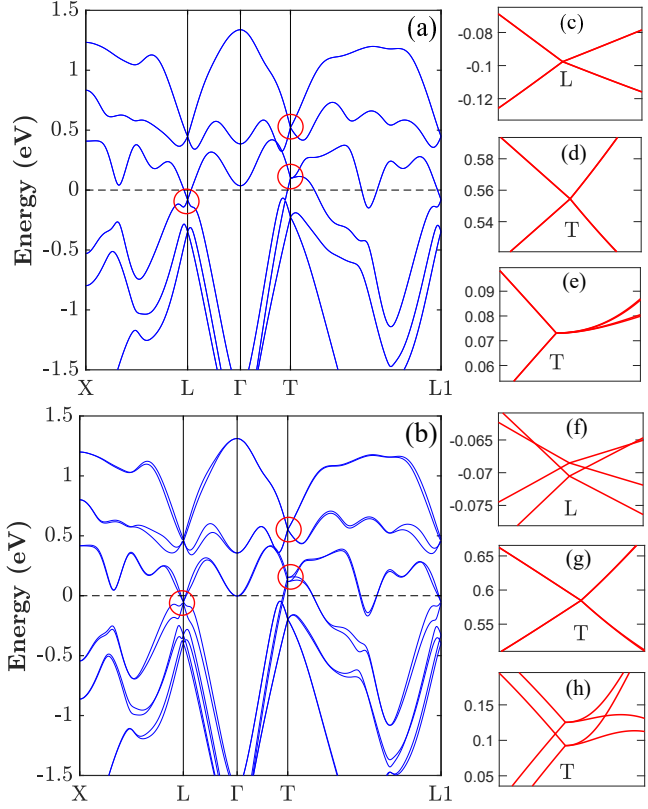}
  \caption{Electronic band structure for (a) the non-ferroelectric phase and (b) the ferroelectric phase where SOC is included. (c)-(h) Zoom-in images of the band crossings at the L and T points marked by red circles in (a) and (b).}\label{fig2:bands}
\end{figure}

\begin{figure}[tbh]
  \centering
  \includegraphics[width=8.5cm]{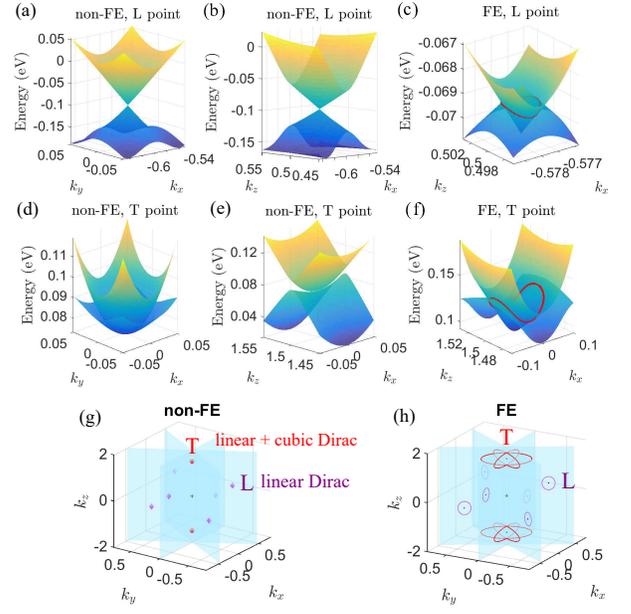}
  \caption{Top and middle panels show the dispersions around the L and T points. (a), (b), (d) and (e) are for the non-ferroelectric phase, while (c) and (f) are for the ferroelectric phase. (g, h) Schematic diagrams showing the band-crossings in the non-ferroelectric (g) and ferroelectric (h) phases.}
  \label{fig3:DPNL}
\end{figure}

The band structure including SOC is of greater interest because the low-energy states mainly arise from the Os-$5d$ orbitals with strong SOC effects. When the SOC is turned on, the band-crossings in the non-ferroelectric \Rc phase evolve from 1D nodal lines into 0D Dirac points, each with four-fold degeneracy. Fig.~\ref{fig2:bands}(a) shows that close to Fermi level, now we have a linear Dirac node at each L point (there are three inequivalent L points in the BZ) [Fig.~\ref{fig2:bands}(c) and Fig.~\ref{fig3:DPNL}(a,b)]. The original eight-fold degeneracy at T splits under SOC into two Dirac points, where one is linear [Fig.~\ref{fig2:bands}(d)] and the other is cubic [Fig.~\ref{fig2:bands}(e) and Fig.~\ref{fig3:DPNL}(d,e)]. Hence, we may call this phase as a multi-type Dirac semimetal.

The cubic Dirac point, around which the dispersion is linear along one direction ($k_z$-axis) and cubic along the other two directions, has been rarely reported in real materials~\cite{QLiu2017}. In previous work, it was believed that such a Dirac point can only appear on a sixfold axis~\cite{YangNaga,QLiu2017} and is possible only for two space groups (No.~176 and No.~192)~\cite{QLiu2017}. Our results clearly demonstrate that it can also occur on a threefold axis in the presence of the additional nonsymmorphic (glide mirror) symmetry. In the following, we shall show that the cubic Dirac point is indeed protected by the glide mirror together with the symmorphic crystal symmetries.

The following points may be noted in connection with symmetry considerations. (1) In the non-ferroelectric phase, the presence of both $\mathcal{T}$ and $\mathcal{P}$ symmetries forces the two-fold spin-degeneracy of each band. Hence the crossing between the bands must be at least 4-fold degenerate (i.e. Dirac type). (2) All states at T and L symmetry points must be degenerate quadruplets. This reason is that Kramers degeneracy at TRIM points requires the eigenvalues of $\mathcal{P}$ to be paired as $(1,1)$ or $(-1,-1)$. In contrast, at T and L,
\begin{equation}
[\mathcal{P}, \My]\neq 0,
\end{equation}
implying that the degenerate states generated by $\My$ will have $\mathcal{P}$ eigenvalues paired as $(1,-1)$, which guarantees the four-fold degeneracy at T and L points. (3) The nature of the Dirac point (linear or cubic) strongly relies on the additional rotational symmetry $\Cz$, which is present at T but not at L, as we will explicitly demonstrate below.

As we already pointed out, the low-energy bands are mainly derived from the Os-$5d$ orbitals. Under trigonal prismatic coordination, the Os $d$-orbitals split into two groups: $A_{1g}$ $(d_{z^2})$ and $E_g$ $\{(d_{x^2-y^2},d_{xy}),(d_{xz}, d_{yz})\}$. Under SOC, these states can be combined into the following spin-orbit-coupled symmetry-adapted basis, keeping in mind that each unit cell contains two Os atoms, one in each sublattice:
\begin{equation}
\phi_{0,s}^\tau=|d_{z^2,s}^\tau\rangle,
\end{equation}
\begin{equation}
\phi_{\pm,s}^\tau=\frac{1}{\sqrt{2}}(\sin\lambda |d_{xz,s}^\tau \pm is d_{yz,s}^\tau\rangle+\cos\lambda |d_{x^2-y^2,s}^\tau\mp  is 2d_{xy,s}^\tau\rangle),
\end{equation}
where $\tau$ labels the two sublattices A and B, $s$ labels the spin, and $\lambda$ is a normalization coefficient.

At the T point, Kramers degeneracy requires that the four basis functions $\phi_{i,s}^\tau$ with a fixed $i=0,\pm$ are coupled together into a quadruplet. Consider first the two quadruplet basis $\Psi_{0/-}=(\phi_{0/-,\uparrow}^A,\phi_{0/+,\downarrow}^A,\phi_{0/-,\uparrow}^B,\phi_{0/+,\downarrow}^B)$, for which the symmetry operations at T take the following representations:
\begin{eqnarray}\label{eq:LDPsymmetry}
\T= \tau_{3}\otimes s_{2}K &,&  \ \
\mathcal{P} = -\tau_{2},\ \nonumber\\
\My = e^{-iq_{z}/2} \tau_{1} \otimes s_{2} &,&\ \
\Cz = \tau_{0}  \otimes e^{ \mp i\frac{\pi}{3} s_{3}}. \end{eqnarray}
Here $\mathbf{\tau}$ and $s$ are Pauli matrices acting on the sublattice and spin spaces, respectively, and $\bm q$ is the wave-vector measured from T. These expressions fix the \kp Hamiltonian at T expressed using $\Gamma$ matrices, which we define here as $\gamma^{1} = \tau_{3}s_{1}$, $\gamma^{2} = \tau_{3}s_{2}$, $\gamma^{3} = \tau_{1}s_{0}$, $\gamma^{4} = \tau_{2}s_{0}$, and $\gamma^{5} = \tau_{3}s_{3}$. Then, to the lowest order, we find that
\begin{eqnarray} \label{eq:LDP}
\mathcal{H}_{\text{T}}^\text{linear} (\bm q) = \alpha_1 (q_{x}\gamma^{1} \pm q_{y}\gamma^{2}) +\alpha_2 q_{z}\gamma^{3} +\alpha_3 q_{z}\gamma^{5},
\end{eqnarray}
where $\alpha$'s are the expansion coefficients, and the sign $\pm$ is for the two basis $\Psi_{0/-}$. This model describes the linear Dirac points at T, which can be viewed as consisting of two Weyl points with Chern numbers 1 and $-1$. As expected, the total Chern number for a closed surface surrounding the Dirac points vanishes.

As for the other quadruplet basis, $\Psi_{+}=(\phi_{+,\uparrow}^A,\phi_{-,\downarrow}^A,\phi_{+,\uparrow}^B,\phi_{-,\downarrow}^B)$, the representations of the symmetry operations are the same as in Eq.~(\ref{eq:LDPsymmetry}) except that $\Cz=-\tau_{0}  \otimes s_{0}$. Due to this different transformation behavior of the basis under $\Cz$, the related effective Hamiltonian is different from (\ref{eq:LDP}), where the diagonal terms proportional to $\tau_0s_0$ have been dropped:
\begin{eqnarray}\label{eq:CDP}
\mathcal{H}_{\text{T}}^\text{cubic} (\bm q)=
\begin{pmatrix}
h_{11} & h_{12} \\
h^{\dagger}_{12} & -h_{11}\\
\end{pmatrix},
\end{eqnarray}
where
\begin{eqnarray*}
h_{11}&=& [c_{1}(q^3_{+}+q^3_{-})+b_1(q_{+}q_{-}q_{z})+a_{1}q_{z}]s_{1} \\ \nonumber
&&+i c_{2}(q^3_{+}-q^3_{-}) s_{2} \\ \nonumber
&&+ [c_{3}(q^3_{+}+q^3_{-})+b_2(q_{+}q_{-}q_{z})+a_{2}q_{z}] s_{3},\\ \nonumber
h_{12} &=& [c_{4}(q^3_{+}+q^3_{-})+b_3(q_{+}q_{-}q_{z})+a_{3}q_{z}] s_{0}.  \nonumber
\end{eqnarray*}
Here $q_\pm=q_x\pm iq_y$, and $a_i$, $b_i$, and $c_i$ are the expansion coefficients. We find that the band-crossing at T described by Eq.~(\ref{eq:CDP}) to be a cubic Dirac point, with cubic dispersion in the $q_x$-$q_y$ plane. The diagonal blocks describe two triple-Weyl fermions with
Chern numbers $\pm 3$. The cubic Dirac point can be viewed as being composed of the two triple-Weyl points.

  We emphasize that when only symmorphic symmetries are considered, cubic Dirac points require the presence of a six-fold axis. ~\cite{YangNaga,QLiu2017}  Our case, however, involves a nonsymmorphic glide mirror, which plays a crucial role in realizing the cubic Dirac point on a three-fold axis. Our analysis indicates that a sufficient condition for realizing a cubic Dirac point is the presence of P, $\T$, a glide mirror, and a c3-axis within the mirror. P, $\T$, and the glide mirror only protect a four-fold degeneracy; the presence of an additional three-fold rotation then yields the cubic in-plane dispersion.

A similar analysis for the L point at $(0,0,\frac{1}{2})$ leads to the following effective Hamiltonian:
\begin{eqnarray}\label{eq:LLDP}
\mathcal{H}_{\text{L}}(\bm q) = \sum\limits_{i=1,3,5} (\beta_{x,i} q_x + \beta_{z,i} q_z)\gamma^{i} + \beta_y q_y \gamma^2.
\end{eqnarray}
Here $\bm q$ is measured from L, and the $\beta$'s are the expansion coefficients. This demonstrates that the crossings at L are linear Dirac points.

\begin{figure}\centering
      \resizebox{8cm}{!}{
              \includegraphics{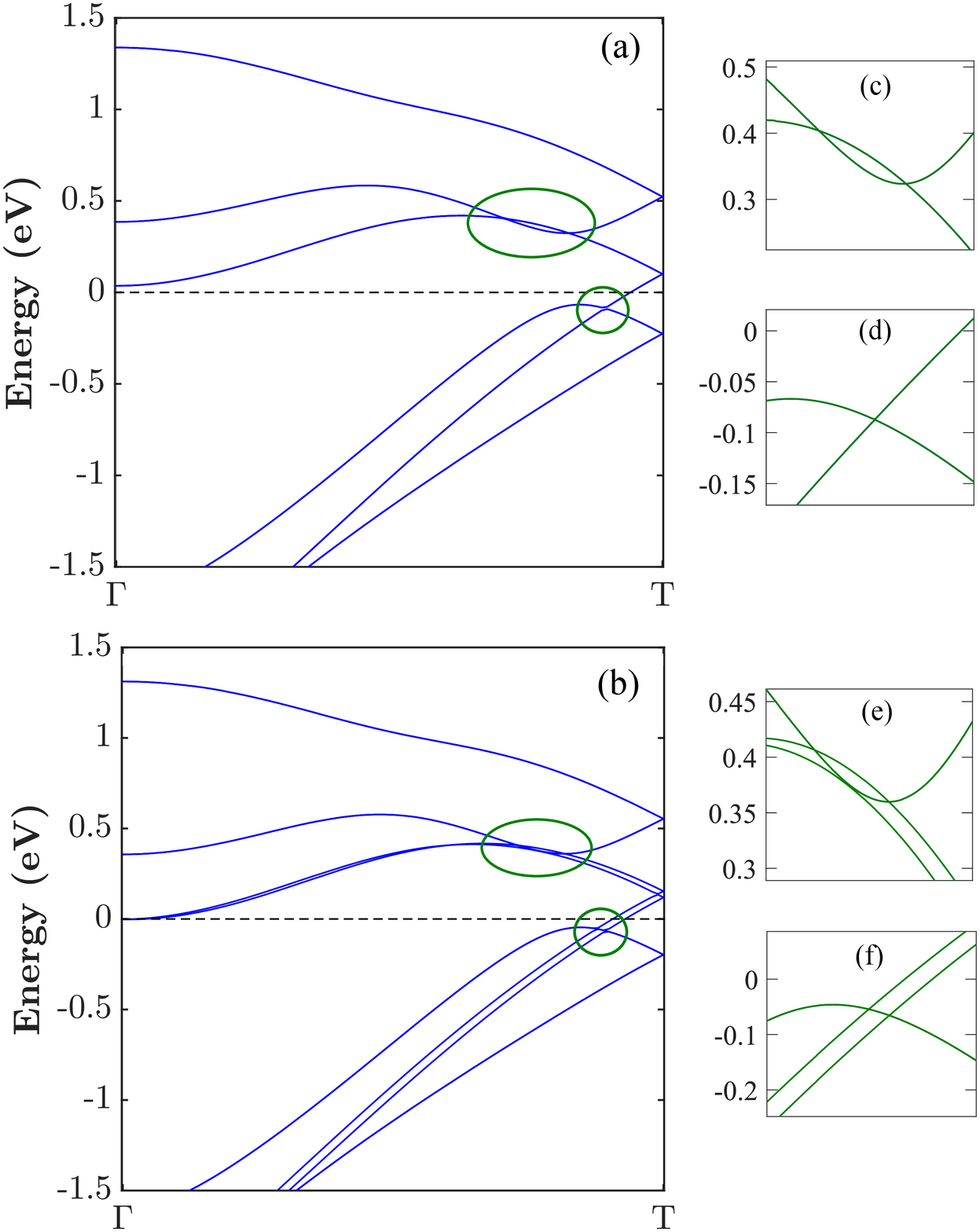}}                				
              \caption{\label{figS2} Accidental band crossing points along the $\Gamma$-T symmetry line. (a) Band structure with SOC for the non-ferroelectric phase. (b) Band structure with SOC for the ferroelectric phase. (c,d) Zoom-in images of the band-crossing points marked by green circles in (a). (e,f) Zoom-in images of the band-crossing points in (b). Each band in (c,d) is two-fold degenerate, so that the crossing points are four-fold degenerate. The crossing points in (e,f) lie between a non-degenerate and a two-fold degenerate band, so that these points are triply degenerate.}
\end{figure}

The linear and cubic Dirac points discussed above are \emph{essential} in the sense that their appearance at the high-symmetry TRIM points T and L is mandated by the symmetry operations of the system. In addition, we note the presence of  accidental Dirac points, which appear in pairs along the primary rotation axis $\Gamma$-T. There are three such pairs around the Fermi level, one of these is seen to be of type-II with an over-tilted dispersion. Zoom-in images of these Dirac points are shown in Figs.~\ref{figS2}(a,c,d).

\subsection{Ferroelectric phase: crossed nodal rings}

In going across the ferroelectric \rc phase transition, the loss of inversion symmetry induces profound changes in the band structure, see Fig.~\ref{fig2:bands}(b). The spin-degeneracy of the bands gets lifted, and the Dirac points in the non-ferroelectric phase become unstable. We find that each linear Dirac point at L develops into a nodal ring on the glide mirror plane, which encloses the L point [Figs.~\ref{fig2:bands}(f) and \ref{fig3:DPNL}(c)]. Also, the cubic Dirac point at T is transformed into three mutually-crossed nodal rings, each lying on a glide mirror plane [Figs.~\ref{fig2:bands}(h) and \ref{fig3:DPNL}(f)]. This topological phase may thus be called a crossed-nodal-ring semimetal. The transformations of these band-crossings across the phase transition are illustrated in Fig.~\ref{fig3:DPNL}(g,h).

The occurrence of the preceding nodal rings in the band structure is also essential in that it is solely determined by symmetry considerations. Note that the glide eigenvalues are $\pm i$ at $\Gamma_{1}\in\{\Gamma, \mathrm{X}\}$, and $\pm 1$ at  $\Gamma_{2}\in \{\mathrm{T},\mathrm{L}\}$. The presence of $\mathcal{T}$ here requires that the Kramers pairs at the above TRIM points carry complex-conjugated eigenvalues, \ie $(i, -i)$ at $\Gamma_1$; and $(1,1)$ or $(-1,-1)$ at $\Gamma_2$. Hence, along any path connecting $\Gamma_1$ and $\Gamma_{2}$ in the glide plane, the evolution of the glide eigenvalues drives a switching of Kramers partners. During this switching, two bands with opposite glide eigenvalues must produce a crossing. As this argument holds for any in-plane path, a nodal loop separating $\Gamma_1$ and $\Gamma_{2}$ must appear. This symmetry analysis highlights the importance of the nonsymmorphic glide mirror in producing these essential band-crossings.

The transformations described above can also be captured in the effective models. For example, at the T point, starting from the model of Eq. \ref{eq:CDP} for the cubic Dirac point, the leading order perturbation $\delta \tau_1 s_2$ when the $\mathcal{P}$ is broken, so that a minimal model may be expressed as
\begin{equation}
\mathcal{H}_{\text{T}}^\text{CNR} (\bm q)=\mathcal{H}_{\text{T}}^\text{cubic} (\bm q)+\delta\tau_1 s_2.
\end{equation}
It is straightforward to verify that this model gives three mutually-crossed nodal rings like the DFT calculations. Similarly, the nodal ring around the L point can be derived from the model of Eq. \ref{eq:LLDP} as follows
\begin{equation}
\mathcal{H}_{\text{L}}^\text{NR} (\bm q)=\mathcal{H}_{\text{L}} (\bm q)+\delta'\tau_1 s_2.
\end{equation}

Note that the linear Dirac points at T are still preserved [Figs.~\ref{fig2:bands}(g)], which can be attributed to the non-commutativity of $\Cz$ and $\My$ in the corresponding basis. The three pairs of accidental Dirac points residing on the primary rotational axis are, however, turned into triply-degenerate nodes as shown in Fig.~\ref{figS2}(b,e,f).

\begin{figure}[tbh]
  \centering
  \includegraphics[width=8.5cm]{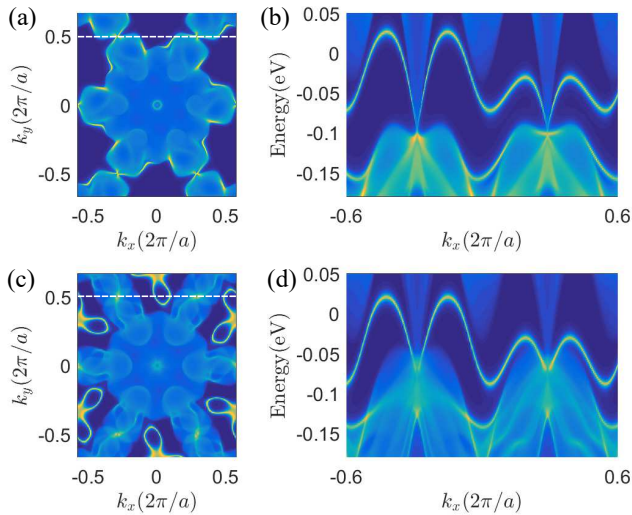}
   \caption{Surface states of \LiO in the non-ferroelectric (upper panel) and ferroelectric (lower panel) phases in the projected spectra of the (111) surface. SOC is included. (a) and (c) Constant energy surfaces at the band-crossing energy at L. (d) and (e) Dispersions along the $k_x$-direction at $k_y=0.5$.} \label{fig4:SS}
\end{figure}

Unlike Weyl points, the essential Dirac points may or may not give rise to nontrivial surface states~\cite{YangNaga}. In Fig.~\ref{fig4:SS}, we plot the surface spectrum of the (111)-surface for both the non-ferroelectric and ferroelectric phases. Both phases are seen to support surface bands connecting the two surface-projected L points.

\section{Conclusion}

We emphasize a number of points in closing. Firstly, the Dirac nodes at the TRIM points and the multiple nodal rings surrounding these points discussed here are \emph{essential} band-crossings in the sense that the presence of these features is solely dictated by the nonsymmorphic space group symmetries (plus $\mathcal{T}$) of the system. Their existence is thus guaranteed as long as these symmetries are preserved. Our analysis of symmetry considerations underlying cubic Dirac points and crossed nodal rings will provide a useful guide in searching for exotic band crossings in realistic material systems more generally.

Secondly, our analysis clearly indicates that \LiO will provide a useful platform for exploring exotic topological phases and their interplay with ferroelectric ordering. The unique advantages of \LiO are as follows. (i) The essential band-crossings are close to the Fermi level, so that the associated topological physics will be reflected in various electronic properties. (ii) The two topological phases that coexist in the same material are connected via a ferroelectric phase transition, which is tunable by varying temperature~\cite{Shi2013}. And, (iii) the material has already been realized experimentally and its ferroelectric phase transition has been observed. The interesting transformations in the band-crossings across the phase transition, which we have predicted here, could be detected by ARPES experiments.

Thirdly, we have identified the first case of an experimentally realized material, which harbors a cubic Dirac point in a 3D material. Such a Dirac point has been predicted previously only in quasi-1D molybdenum monochalcogenide compounds~\cite{QLiu2017}, where an experimental verification is still lacking. Our work offers a new route for realizing cubic Dirac points, extending the range of materials in search of cubic Dirac fermions.

Finally, our study opens a new pathway for exploring a variety of novel effects associated with the various nontrivial band-crossings. For example, a linear Dirac point may exhibit negative magnetoresistance~\cite{Xiong,YGao}, a special magnetic oscillation frequency driven by surface Fermi arcs~\cite{Moll2016}, and an artificial gravity field via strain modulation~\cite{Guan2017}. A cubic Dirac point can exhibit unusual quantum interference contributions to magneto-transport~\cite{HYao}, stronger screening of interactions, and possible presence of continuous quantum phase transitions driven by interactions~\cite{BRoy}. A nodal ring may yield strong anisotropy in electrical transport~\cite{Mullen2015}, unusual optical response~\cite{Ahn} and circular dichroism~\cite{YLiu}, and possible surface magnetism and superconductivity~\cite{Heik}.

\section*{ACKNOWLEDGMENTS}
The work at National University of Singapore was supported by the Singapore National Research Foundation under the NRF fellowship Award No. NRF-NRFF2013-03. The work at Northeastern University was supported by the US Department of Energy (DOE), Office of Science, Basic Energy Sciences grant number DE-FG02-07ER46352, and benefited from Northeastern University's Advanced Scientific Computation Center and the National Energy Research Scientific Computing Center through DOE grant number DE-AC02-05CH11231. FCC and XZ acknowledge support from the National Center for Theoretical Sciences. Work at Singapore University of Technology and Design is supported by Singapore MOE Academic Research Fund Tier 2 (Grant No. MOE2015-T2-2-144). FCC also acknowledges support from the Ministry of Science and Technology of Taiwan under Grants Nos. MOST-104-2112-M-110-002-MY3 and the support under NSYSU-NKMU JOINT RESEARCH PROJECT \#105-P005 and \#106-P005. He is also grateful to the National Center for High-performance Computing for computer time and facilities.

%

\end{document}